\documentclass[runningheads]{llncs}
\usepackage{amssymb,graphicx} 
\begin{document} 
%\frontmatter
\title{{\small University of Iowa Department of Computer Science 
Technical Report 05-13} \\[2ex]
The Poster Session of SSS 2005} 
\author{Brahim Hamid\inst{1} \and Ted Herman (Editor)\inst{2} 
\and Morten Mjelde\inst{3}}
\institute{LaBRI, University of Bordeaux-1, France \and
University of Iowa, USA \and University in Bergen, Norway}
\maketitle

The 2005 Symposium on Self-Stabilizing Systems (SSS 2005) took
place October 26 and 27, 2005, in Barcelona, Spain.  The proceedings
of the Symposium are published in the Springer Lecture Notes on
Computer Science Series, as Volume 3764.  In addition to the 
presentations contained in the proceedings, there were 
five short presentations in a ``poster session'' on 26 October.
Two of the presenters kindly provided short abstracts of their 
presentations (their names are listed as co-authors of this 
technical report).  The five presentations of the poster session
were:
\begin{enumerate}
\item \emph{Self-stabilizing Coloration in Anonymous Planar Networks},
presented by Shing-Tsaan Huang.  This presentation described research
published in \emph{Information Processing Letters}, Volume 95, 
Issue 1, 16 July 2005, Pages 307-312.
\vspace{1ex}
\item \emph{Self-Stabilizing Maximum Matching using Multi-Wave Synchronization}
was presented by Mehmet Hakan Karaata, Kuwait University.
\vspace{1ex}
\item \emph{Electronic Business with Security Modules}, presented by 
Lucia Draque Penso, showed an application of a result later appearing
in Opodis 2005: \emph{Optimal Randomized Omission-Tolerant Uniform 
Consensus in Message Passing Systems}, by 
Felix Freiling, Maurice Herlihy, and Lucia Penso.
\vspace{1ex}
\item \emph{A Formal Model for Snap-Stabilization in Distributed Systems} 
was presented by Brahim Hamid, and a brief abstract of that presentation
appears later in this report.
\vspace{1ex}
\item \emph{Self-stabilizing K-packing and K-domination on tree graphs},
was presented by Morten Mjelde, and a brief abstract of the results
appears later in this report.
\end{enumerate}

%\mainmatter

\title{A Formal Model for Snap-stabilization in Distributed
Systems} 
\author{Brahim Hamid (hamid@labri.fr)} 
\institute{LaBRI, University of Bordeaux-1, F-33405 Talence
Cedex, France} 
\maketitle 

Informally, snap-stabilizing
algorithms ensure that after some transient failures,
the system will automatically recover to reach a correct
configuration in $0$
steps. Therefore, a snap-stabilizing protocol ensures that
the system is in the desirable behavior incessantly. The
most researches about this area are confronted with the
space complexity. In this work, we present a formal method
to deal with snap-stabilization using local computations
and particularly graph rewriting systems: a powerful model
to encode and to prove distributed algorithms. The method
requires the identification of local transient faults. Once
identified, predicates can
be formulated to detect the illegitimate configurations from
the state information and corrective rules can be added to the
algorithm to eliminate each illegitimate configuration. These
corrective rules are formulated such that they do not introduce
any further illegitimate configurations with regards to the
algorithm. Therefore,
the new graph relabeling system is a snap-stabilizing
protocol.
\\ 
\indent A distributed system is modeled by a graph
$G=(V(G),E(G))$ where nodes represent processes and edges
represent bidirectional communication links. The networks are
\textit{asynchronous}, the links are reliable and the process
can fail and recover in a finite time. Every time, each node
and each edge is in some  particular state and this state will
be encoded by a node label or an edge label. According to its
own state and to the states of its neighbors, each node may
decide to realize an elementary {\em computation step}. After
this step, the states of this node, of its neighbors and of
the corresponding edges may have changed according to some
specific {\em computation rules}. Let $L$ be an alphabet and
let $G$ be a graph. We denote by $(G,\lambda)$ a graph $G$
with a relabeling function $\lambda: V(G)\cup E(G)\rightarrow
L$. A graph relabeling system is a triple $\Re=(L,I,P)$ where
$L$ is a set of labels, $I$ is a subset of $L$ called the set
of initial labels and $P$ a finite set of relabeling rules.
\\ \indent Local configurations will be defined on balls
of radius 1 denoted by $B$ (the corresponding node and the
set of its neighbors). For a labeled graph $(G,\lambda)$,
we say that a local configuration
$f=(B_f,\lambda_f)$ is illegitimate for $(G,\lambda)$, if
there is no ball (neither sub-ball) of radius $1$ in $G$
which has the same labeling as $f$. A graph relabeling
system with illegitimate configuration is a quadruple
$\Re_{ic}=(L,I,P,\cal F)$ where $\cal F$ is a set of
illegitimate configurations. A local snap-stabilizing graph
relabeling system is denoted by $\Re_{sn}=(L,\cal P)$ 
where $\cal P$ a finite set of relabeling rules composed
of the set of relabeling rules $P$ and some
correction rules $P_c$. The rules of the set  $P_c$
are introduced in order to eliminate the illegitimate
configurations. Therefore, these rules delay the safe
computation which is a computation without fault of
components. Our approach is easy to understand and its
translation from the initial algorithm requires little
changes and can be applied in practical applications as a
generic and automatic method to deal with transient faults
in distributed systems.

\author{Morten Mjelde \and Fredrik Manne}
\title{Self-stabilizing $K$-Packing and $K$-Domination on tree graphs}
\institute{mortenm@ii.uib.no, fredrik.manne@ii.uib.no \\ 
Department of Informatics, University in Bergen}

\maketitle

The self-stabilizing algorithms presented
here solves the problems of maximum $K$-packing and minimum
$K$-domination on a tree graph in $O(n^{3})$ moves (assuming
an adversarial daemon). Previously known algorithms were
able to solve maximal 2-packing and minimal $k$-domination
(as opposed to maximum and minimum), both on general graphs,
and both in exponential time. The algorithms presented here
solves the problems by using two passes through the tree. The
first one finds the size of the optimal solution for the
entire tree, and the second pass finds the actual solution.

$K$-packing is defined as follows: Assume a
graph $G=(V,E)$. A $K$-packing algorithm selects a subset $S$
(black vertices) of the vertices $V$ such that for every pair
$(v,u) \in S$ the shortest distance between them is always
greater than $K$ (that is, there has to be $K$ or more vertices
between $v$ and $u$). Maximum $K$-packing means that no subset
of $V$ that is a legal solution has a large cardinality than
$S$. Maximum $K$-packing is NP-Hard for a general graph.

The self-stabilizing algorithm presented here
solves maximum $K$-packing on a tree graph. It is assumed
that every vertex knows which of its neighbours is closest to
the root (ie. its parent) (there are however self-stabilizing
algorithms that can find this without effecting the asymptotic
running time). The algorithm solves the problem in two phases:
Phase 1 finds the optimal solution for the subtree $T[v]$
(the subtree with the vertex $v$ as the root) for every $v
\in V$. The calculation is performed using information about
the optimal solution found in $v$'s children. Phase 2 sets
appropriate vertices to black based on what was found in
Phase 1. The algorithm can be shown to stabilize in no more
then $O(n^{3})$ moves.

For each $v \in V$, there exists a table
$M_{v}[0:K]$. When the table has been computed, each index $i$
holds the size of the optimal solution for the subtree $T[v]$
assuming that there are at least $i$ levels of white vertices
(ie. vertices not belonging to $S$) below $v$ (including $v$
itself). $i=0$ implies that $v$ can be black, $i=1$ implies
that $v$ is white, but a child may be black and so on. The
table for a vertex $v$ is computed using the tables $M_{c}[]$
for every child $c$ of $v$. When the table $M_{r}[]$ for
the root $r$ is computed, position $0$ shows the size of the
optimal solution for the entire tree.

Phase 2 finds the actual vertices belonging
to $S$ by backtracking from Phase 1. It sets a vertex $v$
to black depending on which position in the table $M_{v}[]$
was found to house the optimal solution for the subtree $T[v]$.

$K$-domination is similar to $K$-packing, and
is defined as follows: A subset $S$ of the vertices in $G$
is selected such that for every vertex $v \in V$ there exists
at least one vertex $s \in S$ where the distance between $v$
and $s$ is less than or equal to $K$. Minimum $K$-domination
means that no subset of $V$ that is a legal solution has
a smaller cardinality than $S$. Minimum $K$-domination is
NP-complete on a general graph. Because of the similarities
with $K$-packing, $K$-domination can be solved by using much
the same algorithm as described above. The number of moves
will be the same in both cases.

\end{document}